\documentclass[amsfonts,notitlepage,prl,aps,twocolumn]{revtex4-1}

\usepackage{amsmath}
\usepackage{amssymb}
\usepackage{color}
\usepackage{datetime}
\usepackage{braket}
\usepackage{siunitx}
\usepackage{graphicx}

\newcommand{\E}{\boldsymbol{\mathcal E}}
\newcommand{\e}{\boldsymbol \epsilon}

\begin{document}
\title{Chiral nanophotonic waveguide interface based on spin-orbit coupling of light}

\author{Jan Petersen}
\author{J\"urgen Volz}
\author{Arno Rauschenbeutel}
\affiliation{Vienna Center for Quantum Science and Technology, TU Wien -- Atominstitut, Stadionallee 2, 1020 Vienna, Austria}



\begin{abstract}
Controlling the flow of light by means of nanophotonic waveguides has the potential of transforming integrated information processing much in the same way that conventional glass fibers have revolutionized global communication. Owing to the strong transverse confinement of the light, such waveguides give rise to a coupling between the internal spin of the guided photons and their orbital angular momentum.
Here, we employ this spin-orbit coupling of light to break the mirror symmetry of the scattering of light by a single gold nanoparticle on the surface of a nanophotonic waveguide. We thereby realize a chiral waveguide coupler in which the handedness of the incident light determines the direction of propagation in the waveguide. Using this effect, we control the directionality of the scattering process and direct up to 94\% of the incoupled light into a given direction 
This enables novel ways for controlling and manipulating light in optical waveguides and nanophotonic structures as well as for the design of integrated optical sensors.
\end{abstract}
\maketitle
 
The development of integrated electronic circuits laid the foundations for the information age which fundamentally changed modern society. During the last decades, a transition from electronic to photonic information transfer has taken place and, nowadays, integrated nanophotonic circuits and waveguides promise to partially replace their electronic counterparts and even enable radically new functionalities\cite{Gramotnev2010,Benson2011,Tong2012}. The strong confinement of light provided by such waveguides  leads to significant intensity gradients on the wavelength scale. In this strongly non-paraxial regime, spin and orbital angular momentum of the guided light are no longer independent physical quantities but are coupled\cite{Bliokh2012,Bliokh2014}. In particular, the spin depends on the position in the transverse plane and on the direction of propagation of light in the waveguide -- an effect referred to as spin-orbit coupling of light. Spin-orbit coupling holds great promises for the investigation of a large range of physical systems, comprising phenomena such as the spin-Hall effect\cite{Klitzing1980,Murakami2003,Koenig2007} and extraordinary momentum states\cite{Bliokh2014}, and has been observed in the case of total internal reflection in a prism\cite{Hosten2008} as well as in plasmonic systems\cite{Bliokh2008,Gorodetski2012,Rodriguez2013,Lin2013}. Recently, it has been demonstrated in a cavity-quantum electrodynamics setup that spin-orbit coupling fundamentally modifies the interaction between a single atom and the resonator field\cite{Junge2013}.

\begin{figure}
\centerline{\includegraphics[width=0.9\columnwidth]{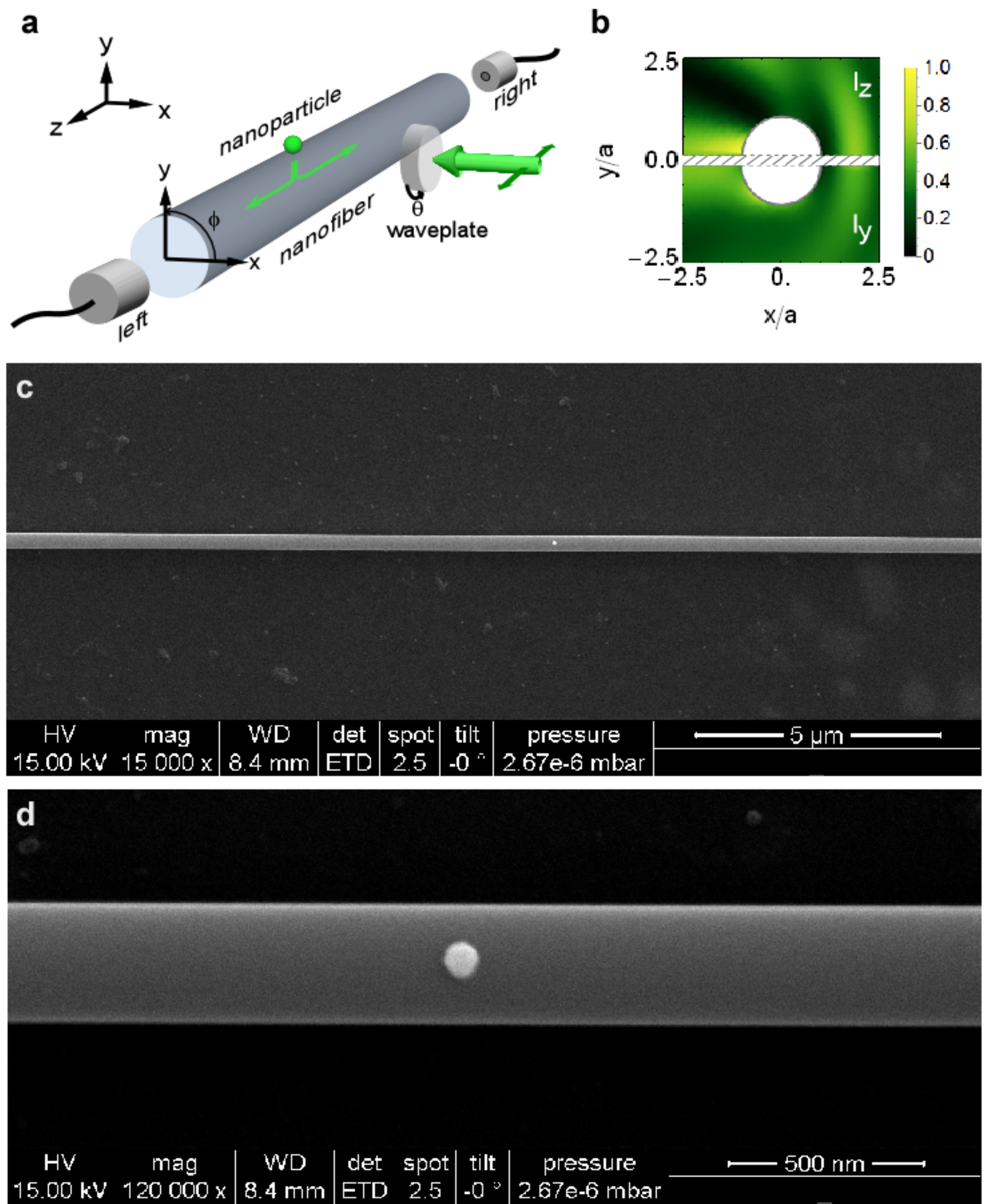}}
\caption{\textbf{Experimental setup. a} A single nanoparticle is deposited on a silica nanofiber and illuminated with a laser beam that propagates in the negative $x$-direction. The polarization of the light can be set with a quarter wave plate.  The fiber can be rotated around the $z$-axis by the angle $\phi$, which amounts to changing the azimuthal position of the particle. Here, $\phi=90\si{\degree}$ corresponds to the case where the nanoparticle is on the top of the fiber. The light scattered into the nanofiber is detected using single photon counting modules at each output port (left and right) of the fiber. \textbf{b}  The presence of the nanofiber modifies the intensity distribution of the incident light field. The relative intensity distributions for an incident field polarized along $y$- and $z$-axis are shown. \textbf{c,d} Scanning electron microscope images of the nanofiber and the single nanoparticle used in our experiments. From the images, we determine diameters of $2a=(315\pm3)\,\si{nm}$ for the fiber and $2r=(90\pm3)\,\si{nm}$ for the nanoparticle.}
\label{FigSetup}
\end{figure}

In the case of vacuum-clad dielectric waveguides, evanescent fields typically arise in the vicinity of the surface of the structure and allow one to locally probe the guided fields and to interface them with micro- and nanoscopic emitters. Due to  spin-orbit coupling, these evanescent fields exhibit a locally varying ellipticity which stems from a longitudinal polarization component that points in the direction of propagation of the light and that oscillates in quadrature with respect to the transversal polarization components\cite{Snyder83,LeKien04b}. Surprisingly and in contrast to paraxial light fields, the corresponding photon spin is in general not parallel or antiparallel to the propagation direction of the guided light. In special cases that will be discussed in more detail below, it can even be perpendicular to the propagation direction and antiparallel to the orbital angular momentum\cite{Bliokh2012,Bliokh2014}.

In our work, we experimentally demonstrate that spin-orbit coupling in a dielectric nanophotonic waveguide drastically changes the scattering characteristics of a nanoscale particle located in the evanescent field. In free space, point-like scatterers exhibit a dipolar emission pattern\cite{Scully97,Jackson}, with a point symmetric scattering rate. This means that for any given optical axis an equal amount of light is scattered into opposite directions. Here, we demonstrate that spin-orbit coupling breaks this symmetry. In particular, when light is scattered by the particle into the waveguide modes, the amount of light that is coupled into a given direction of the waveguide can significantly exceed the power that propagates in the opposite direction. Using this effect, we realize a nanophotonic waveguide coupler in which the polarization of the incident light determines the direction of propagation in the waveguide and direct up to 94\% of the incoupled light into a given direction. 
We expect spin-orbit coupling-enabled devices like the one demonstrated here to have an important impact on controlling and manipulating optical signals in waveguides and nanophotonic structures for future applications.

In our experiment, we employ an air-clad silica nanofiber as an optical waveguide. We position a single spherical gold nanoparticle on its surface, illuminate the particle with a freely propagating, focussed paraxial laser beam from the side (see Fig.~\ref{FigSetup}\textbf{a}), and characterize the scattering properties of the particle into the optical waveguide.
The emission rate of the particle into a given nanofiber eigenmode is proportional to  $|\textbf{d}\cdot \e^*|^2$ with the induced electric dipole moment $\textbf{d}$ of the particle and the profile function $\e$ of the electric part of the fiber mode. For spherical scatterers, the dipole moment is given by $\textbf{d} =\alpha\cdot\E_\textrm{exc}$, where $\alpha$ is the complex polarizability and $\E_\textrm{exc}$ is the positive frequency envelope of the excitation field which is related to the real value of the electric field by $\mathbf{E_\textrm{exc}}=1/2(\E_\textrm{exc}\exp( -i\omega t)+c.c.)$. Here, $\omega/2\pi$ is the frequency of the light and $c.c.$ the complex conjugate. The total power of the light scattered into a given fiber mode is then given by
\begin{equation}
I_{\textrm{scat}}\propto |\textbf{d}\cdot\e^*(r,\phi)|^2= |\alpha\E_\textrm{exc} \cdot\e^*(r,\phi)|^2,
\label{eqn1}
\end{equation}
where $(r,\phi)$ denotes the position of the scatterer in the nanofiber transverse plane. As a consequence, the emission rate is directly proportional to the overlap between the field of the excitation light and the fiber mode at the particle's position.

\begin{figure}
\centerline{\includegraphics[width=1.05\columnwidth]{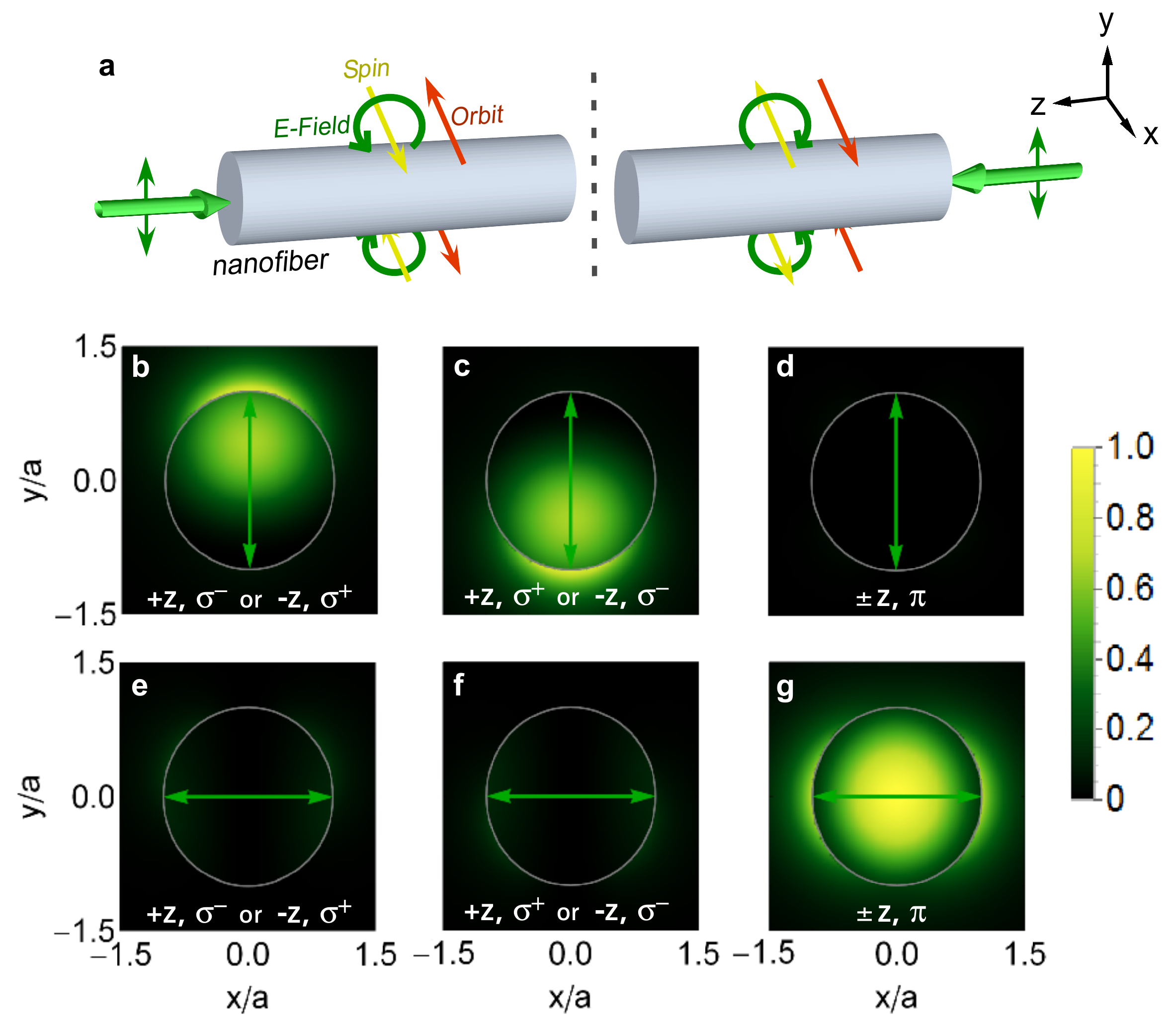}}
\caption{\textbf{Spin-orbit coupling in optical nanofibers.} \textbf{a} When the guided light is quasi linearly polarized along the $y$-axis, longitudinal polarization components occur.  For light traveling in $+z$-direction, this leads to nearly circular $\sigma^-$ ($\sigma^+$) polarization  on the top (bottom) of the fiber, see circular green arrows.  For light propagating in $-z$ direction, $\sigma^-$ and $\sigma^+$ are interchanged. As a consequence, the spin angular momentum of the light (yellow arrows) is oriented perpendicular to the propagation direction and anti-parallel to the orbital angular momentum (red arrows).
\textbf{b} Overlap between the $\textrm{HE}^+_{11,y}$ mode ($\textrm{HE}^-_{11,y}$ mode) and $\sigma^-$ ($\sigma^+$) polarization. \textbf{c} Overlap between the $\textrm{HE}^+_{11,y}$ mode ($\textrm{HE}^-_{11,y}$ mode) and $\sigma^+$ ($\sigma^-$) polarization. \textbf{d} Overlap between the $\textrm{HE}^\pm_{11,y}$ modes and $\pi$ polarization.
\textbf{e-g} Same as \textbf{b-d} but for the fiber modes $\textrm{HE}^\pm_{11,x}$.
 The values are calculated for our experimental parameters (nanofiber diameter: $2a=315\,\si{nm}$, optical wavelength: $\lambda=532\,\si{nm}$) and the fiber mode profile functions are normalized such that $|\e^\pm_{HE,i}(x=y=0)|^{2}=1$.}
\label{FigModes}
\end{figure}

\begin{figure*}
\centerline{\includegraphics[width=1.7\columnwidth]{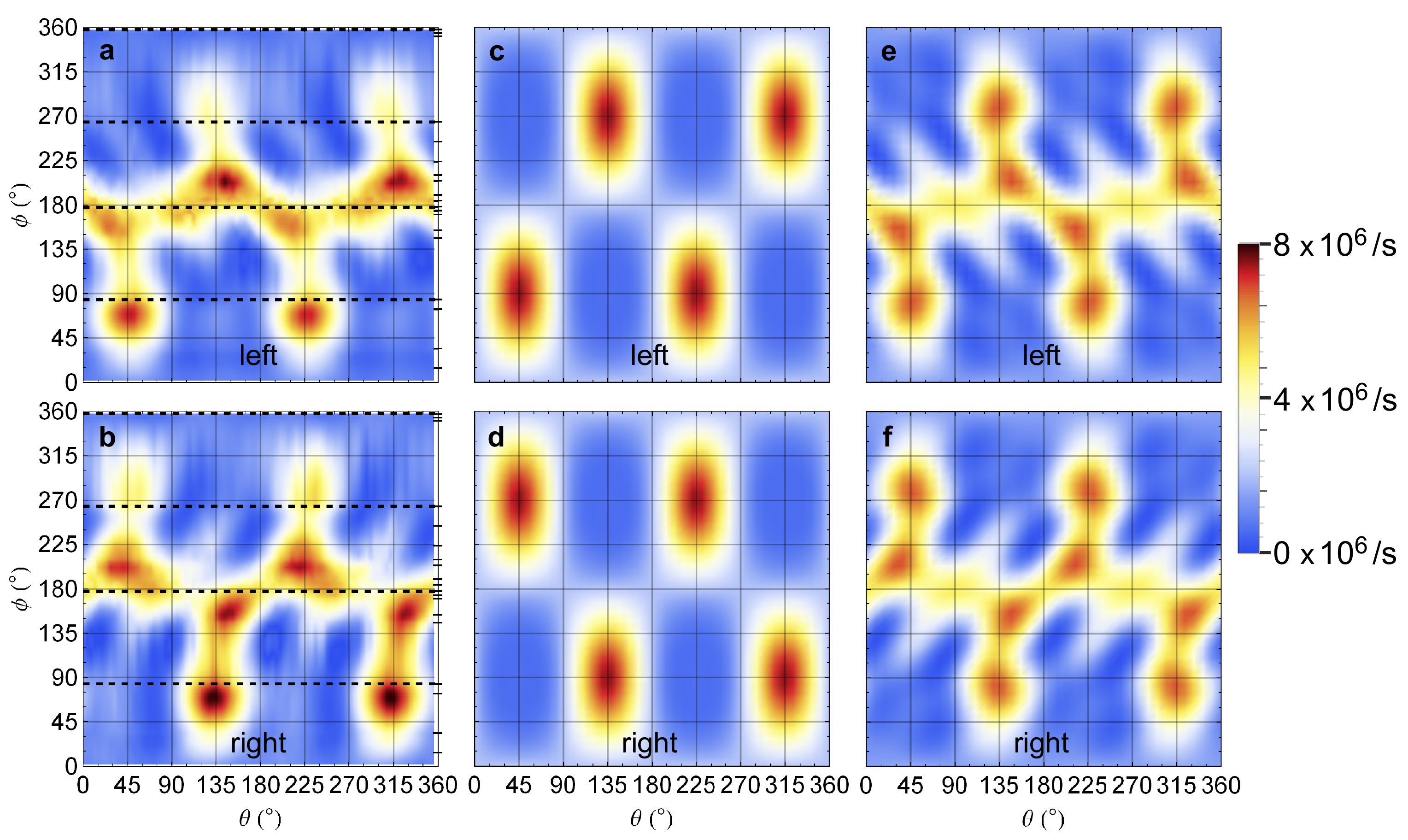}}
\caption{\textbf{Chiral waveguide coupling, experiment and theory.} 
\textbf{a} (\textbf{b}) Measured photon flux of the light scattered into the left (right) direction as a function of the azimuthal position of the nanoparticle $\phi$ and the polarization of the excitation light field set by the rotation angle $\theta$ of the quarter wave plate. The ticks on the right side of the panels mark the azimuthal positions for which data has been acquired with a stepsize of $\theta$ of $5^\circ$. The data is interpolated in between the measured points.
The dashed lines indicate the datasets plotted in Fig.~\ref{FigResults1}. \textbf{c},\textbf{d} (\textbf{e},\textbf{f}) Theoretical prediction for the photon fluxes when neglecting (including) the effect of the nanofiber on the incident light field using two free parameters that are obtained from a fit of \textbf{e} and \textbf{f} to the data in \textbf{a} and \textbf{b} (see methods).}
\label{FigResults2}
\end{figure*}

For a single-mode nanofiber, all guided light fields  can be decomposed into the quasi linearly polarized fiber eigenmodes\cite{Snyder83,LeKien04b} $\textrm{HE}^\pm_{11,x}$ and $\textrm{HE}^\pm_{11,y}$, where the $z$-axis coincides with the nanofiber axis and the $\pm$ sign indicates the propagation direction of the light in the fiber ($\pm z$). We choose $\textrm{HE}^\pm_{11,x}$ and $\textrm{HE}^\pm_{11,y}$ such that their main polarization component points along the $x$-direction ($\phi=0^\circ$) and the $y$-direction ($\phi=90^\circ$), respectively. 
Figure~\ref{FigModes} shows the overlap of the profile functions $\e^\pm_{\textrm{HE},x}$ and $\e^\pm_{\textrm{HE},y}$ of the electric part of the fiber modes (see methods) with circular $\sigma^\pm=(i\mathbf{e}_z\pm \mathbf{e}_y)/\sqrt{2}$ and linear $\pi=\mathbf{e}_x$ polarization as a function of the position in the fiber transverse plane for the parameters used in our experiment, where we have chosen $x$ as the quantization axis. Here, $\mathbf{e}_{x,y,z}$ are the unit vectors along the corresponding axes. From Fig.~\ref{FigModes} it is apparent that the local polarization depends both on the position in the fiber transverse plane and on the direction of propagation of the mode. This is a clear signature of spin-orbit coupling of the nanofiber guided light. We now consider the situation where the particle is located at the top ($\phi=90^\circ$) of the nanofiber. At this position, the overlap between  the $\textrm{HE}^+_{11,y}$ mode ($\textrm{HE}^-_{11,y}$ mode) and $\sigma^-$ polarization is maximal (minimal) and reaches 93\% (7\%). This means that the light is nearly perfectly circularly polarized. Since the emission probability of the particle into the fiber is directly proportional to this overlap, this results in a strong asymmetry of the scattering into the left ($+z$) and right ($-z$) direction of the fiber which can be tuned by the polarization of the incident light field and the  position of the nanoparticle in the fiber transverse plane. In particular, the asymmetry reverses when switching the polarization of the excitation light from $\sigma^-$ to $\sigma^+$ or when changing the position of the particle from $(x,y)$ to $(x,-y)$.

In order to experimentally investigate this directional scattering, we prepare a tapered optical fiber (TOF) with a nanofiber waist\cite{Brambilla2010} (diameter: $2a=(315\pm3)\,\si{nm}$). The TOF enables almost lossless coupling of light fields that are guided in a standard optical fiber into and out of
the nanofiber section. We position a single spherical gold nanoparticle (diameter: $2r=(90\pm3)\,\si{nm}$) on the nanofiber surface which, to a good approximation, acts as a polarization maintaining scatterer\cite{bohren08,Myroshnychenko2008}. This is accomplished by approaching the fiber with a droplet containing a watery suspension of gold nanoparticles which touches the fiber over a length of approximately $100\,\mu$m. By monitoring the transmission through the fiber, we detect the successful deposition of a single gold nanoparticle via a decrease in fiber transmission of around $9\%$ at a wavelength of $532\,\si{nm}$ (see methods). The position of the particle along the fiber is determined by sending light through the nanofiber and by recording the light scattered by the nanoparticle with a microscope. 
 We illuminate the particle with a laser beam that propagates in the negative $x$-direction (see Fig.~\ref{FigSetup}\textbf{a}). It has a wavelength of $532\,\si{nm}$, close to the measured resonance of the nanoparticle with a full width at half maximum of about 50~nm around the central wavelength of $530\,\si{nm}$. The nanofiber is fixed on a mount which can be rotated around the $z$-axis. Due to the cylindrical symmetry of the fiber, this rotation amounts to changing the azimuthal position $\phi$ of the nanoparticle around the fiber (see Fig.~\ref{FigSetup}\textbf{a}). The polarization of the incident light field
 is set by means of a quarter wave plate. The angle $\theta$ between its optical axis and the $y$-axis can be adjusted at will. Before passing through the wave plate, the polarization of the light is aligned along $z$. Thus, we can set the polarization to linear along $z$ ($\theta=0^\circ$, $90^\circ$) and circular, i.e., $\sigma^-$ ($\theta=45^\circ$) or $\sigma^+$ ($\theta=135^\circ$). For intermediate angles the polarization is elliptical with the major axis along $z$.
The excitation laser beam has a waist radius of around $w=150\,\mu$m at the position of the nanoparticle, thereby assuring a fairly homogenous spatial intensity distribution with negligible longitudinal polarization components. In order to detect the light that is scattered into the nanofiber, we use a single photon counting module (SPCM) at each output port of the TOF.
After completion of all measurements, we check that only a single nanoparticle was present,
by analyzing the fiber surface with a scanning electron microscope (see Fig.~\ref{FigSetup}\textbf{c} and \textbf{d}). This also allows us to measure the diameter of the fiber and of the nanoparticle.

Figure~\ref{FigResults2} shows the measured photon flux at the left (\textbf{a}) and right (\textbf{b}) fiber output as a function of the azimutahl position of the  nanoparticle and of the polarization state of the excitation light field (fixed by the wave plate angle $\theta$). Panels \textbf{c} and \textbf{d} show the theoretical predictions calculated according to equation~(\ref{eqn1}) under the assumption that the polarization and intensity distribution of the incident light field are not modified by the presence of the optical fiber (see methods). We find qualitative agreement between our measurement results and the theoretical predictions. In particular, we observe the expected maximum of the left--right asymmetry for the case of circular input polarization with the particle located at the top or at the bottom of the fiber. However, scattering and refraction of the excitation light field by the nanofiber leads to a significant modification of the polarization and intensity of the field close to the nanofiber surface, see Fig.~\ref{FigSetup}\textbf{b}\cite{barber90}. When including these effects, we obtain the theoretical predictions shown in panels \textbf{e} and \textbf{f}, where we used two fit parameters: the angular offset of the nanoparticle from the expected deposition position of $\phi=90^\circ$ and the amplitude of the photon flux detected by the SPCMs (see methods). This model agrees well with the measured data, the main differences to the simple model being an increase  of the scattering rate around $\phi=180^\circ$ due the focusing of the incident light field by the fiber and the emergence of a shadow region around $\phi=120^\circ$ and $\phi=240^\circ$ with a concomitant decrease in the scattering rate.

\begin{figure}
\centerline{\includegraphics[width=1.05\columnwidth]{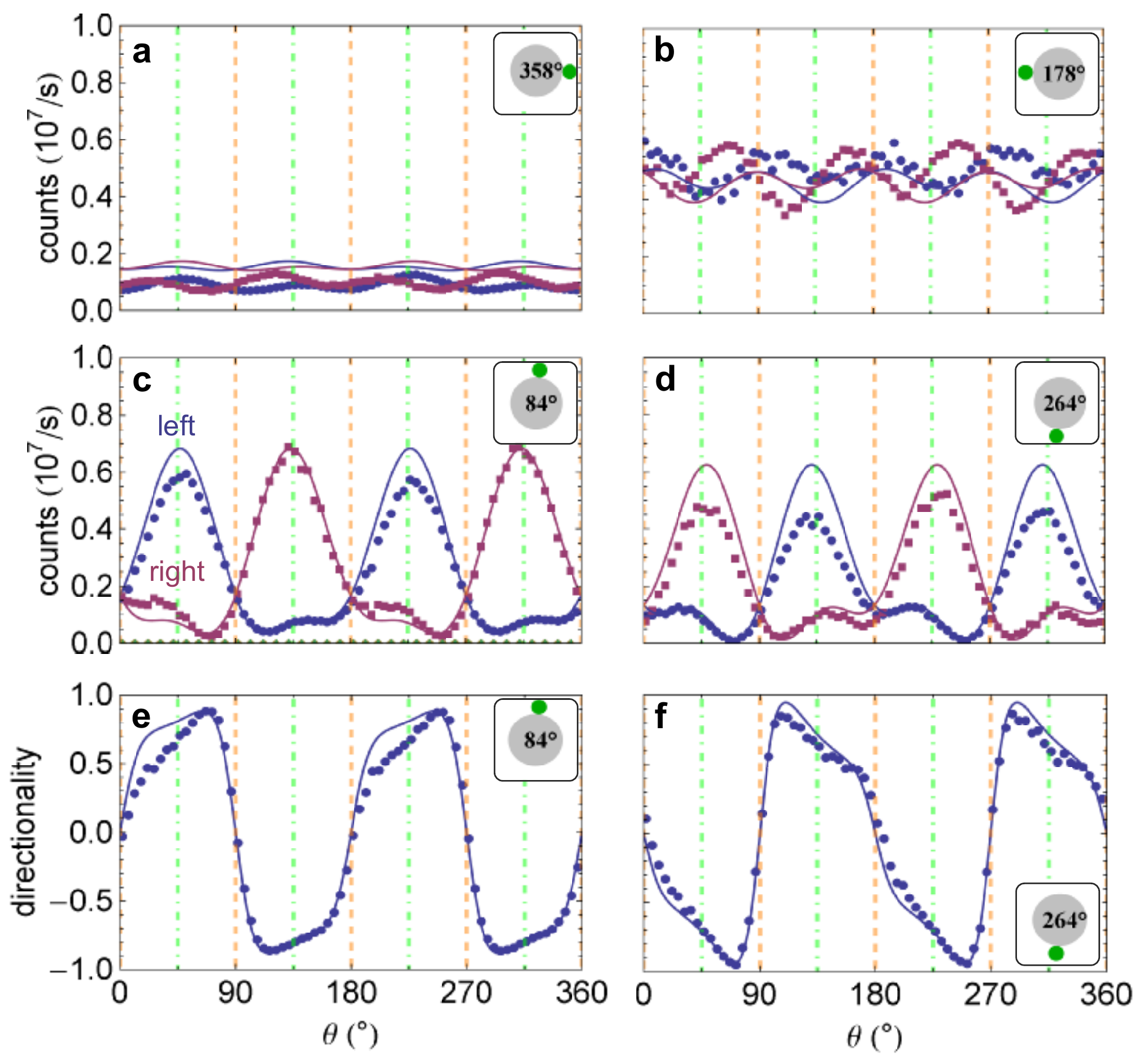}}
\caption{\textbf{Directionality of the scattering process.} \textbf{a}-\textbf{d} Measured photon fluxes at the left (blue circles) and right (red squares) fiber output port as a function of the rotation angle $\theta$ of the quarter wave plate. Here, $\theta=0\si{\degree},90\si{\degree},...$ corresponds to linear polarization along the fiber axis (dashed orange lines) and $\theta=45\si{\degree},225\si{\degree}$ ($\theta=135\si{\degree},315\si{\degree}$) corresponds to $\sigma^-$ ($\sigma^+$) polarization of the incident light field (dash-dotted green lines). \textbf{a}-\textbf{d} correspond to four different azimuthal positions ($\phi=358\si{\degree}$, $178\si{\degree}$, $84\si{\degree}$ and $264\si{\degree}$) of the nanoparticle (green dot) around the fiber (gray disk), as indicated in the insets. The solid lines are the predictions of our theoretical model. The statistical error bars are too small to be visible in the plot. Figure~\textbf{c} also shows exemplarily the measured photon fluxes to the left (yellow diamonds) and right (green triangles) for the nanofiber without the nanoparticle, which are close to zero. \textbf{e},\textbf{f} Directionality $\mathcal{D}$ of the scattering process into the fiber for the data in \textbf{c} and \textbf{d}. }
\label{FigResults1}
\end{figure}

For closer comparison, Fig.~\ref{FigResults1}\textbf{a}-\textbf{d} shows the polarization dependence of the measured photon flux in the fiber for selected azimuthal positions of the nanoparticle together with the theoretical prediction. For the cases of the nanoparticle positioned near the top and the bottom of the nanofiber, we also plot the directionality
\begin{equation}
\mathcal{D}=\frac{c_+-c_-}{c_++c_-}
\label{eqn2}
\end{equation}
of the scattering process together with the theoretical prediction, see panels \textbf{e} and \textbf{f}. Here, $c_+$ and $c_-$ are the photon fluxes detected on the left and right detector, respectively. We observe a maximum directionality of $\mathcal{D}=0.88$ ($\mathcal{D}=0.95$) for a particle near the top (bottom) of the fiber which corresponds to a ratio of $16$:$1$ ($40$:$1$) between the photon flux scattered to the left and right (right and left). When the particle is located at a position near the side of the fiber, the overlap of the fiber eigenmodes with any polarization of the excitation light is independent of the propagation direction and zero directionality is expected. In the experiment, we indeed observe only a small variation with the incident polarization, see Fig.~\ref{FigResults1}\textbf{a} and \textbf{b}. The residual modulation is most probably due to the small angular deviation of the nanoparticle position from the ideal point.

In summary, using a gold nanoparticle on a nanofiber we realized a chiral waveguide coupler, which enables a high-contrast channeling of light into a given direction of the waveguide. The underlying physical mechanism enabling this device is spin-orbit coupling of light which universally occurs in light fields that are strongly confined in the transversal direction. Our method is thus highly versatile and may find application in various scenarios of nanophotonic systems. There is no fundamental limit to the directionality of the mechanism. As an example, the quasi linearly polarized guided modes of our silica nanofiber locally exhibit perfect circular polarization inside the waveguide. Ideally, the scattering of a particle that is placed at such a position and that is excited with circularly polarized light will thus be perfectly suppressed for one direction of the waveguide and will only couple into the other.
Apart from its usefulness for optical signal processing and routing, our scheme enables novel nanophotonic sensing applications, that allow one to detect, e.g., scatterers with 
 an intrinsic polarization asymmetry, such as chiral molecules.

\bibliography{channeling}

\clearpage

\section*{Methods}
\footnotesize
\subsection{Polarization of nanofiber modes}
Nanofiber waveguides have a cut-off frequency below which only the fundamental $\rm{HE}_{11}$ mode can propagate. The single-mode condition  is given by 
$V=k_0a\sqrt{n_1^2-n_2^2}<2.405$, where $k_0$ is the wave vector of the light in vacuum, $a$ is the nanofiber radius and $n_1$ and $n_2$ are the refractive indices of the medium of the fiber and the surrounding medium, respectively. Below the cut-off, any fiber guided field can be expressed as a superposition of the two quasi linearly polarized fiber modes $\rm{HE}^\pm_{11,x}$ and  $\rm{HE}^\pm_{11,y}$ where the profile functions $\e$ of the electric part are given by\cite{LeKien04b, LeKien13c}
\begin{widetext}
\begin{align}
 \epsilon_x & = A \frac{\beta}{2h}  \left((1-s)J_0(hr) \cos(\varphi_0)-(1+s)J_2(hr)\cos(2\phi-\varphi_0)\right)e^{\pm i \beta z} \label{eqnFieldsStart} \\
 \epsilon_y & = A \frac{\beta}{2h}  \left((1-s)J_0(hr) \sin(\varphi_0)-(1+s)J_2(hr)\sin(2\phi-\varphi_0)\right)e^{\pm i \beta z} \\
 \epsilon_z & = \pm iA J_1(hr) \cos(\phi-\varphi_0)e^{\pm i \beta z}
\end{align}
for the field inside the fiber ($r<a$) and 
\begin{align}
 \epsilon_x & = A \frac{\beta}{2h} \frac{J_1(ha)}{K_1(qa)} \left((1-s)K_0(qr) \cos(\varphi_0)+(1+s)K_2(qr)\cos(2\phi-\varphi_0)\right)e^{\pm i \beta z} \\
 \epsilon_y & = A \frac{\beta}{2h} \frac{J_1(ha)}{K_1(qa)} \left((1-s)K_0(qr) \sin(\varphi_0)+(1+s)K_2(qr)\sin(2\phi-\varphi_0)\right)e^{\pm i \beta z} \\
 \epsilon_z & = \pm iA \frac{J_1(ha)}{K_1(qa)}K_1(qr) \cos(\phi-\varphi_0)e^{\pm i \beta z} \label{eqnFieldsEnd},
\end{align}
\end{widetext}
for the field outside the fiber ($r>a$), where the fiber is aligned along the $z$-direction. The propagation constant $\beta$ ($\beta>0$) is determined by the fiber eigenvalue equation\cite{Snyder83} and $s$ is defined as $s=(1/(h^2 a^2)+1/(q^2 a^2))/[J_{1}'(h a)/(h a J_{1}(h a))+K_{1}'(q a)/(q a K_{1}(q a))]$. The parameters $h=(n_{1}^2 k_0^2-\beta^2)^{1/2}$ and $q=(\beta^2-n_{2}^2 k_0^2)^{1/2}$ characterize the fields in- and outside of the fiber, respectively. $J_i$ and $J_i'$ ($K_i$ and $K_i'$) are the Bessel functions of the first kind (modified Bessel functions of the second kind) of order $i$ and their respective derivative  and $A$ is a normalization constant. The  $\pm$ sign indicates the propagation direction and $\varphi_0$ defines the orientation of the principal polarization axis, where $\varphi_0=0\si{\degree}$ ($\varphi_0=90\si{\degree}$) corresponds to the fiber mode $\rm{HE}^\pm_{11,x}$ ($\rm{HE}^\pm_{11,y}$), quasi linearly polarized along the $x$- ($y$-) direction. 

\subsection{Longitudinal polarization components}
In contrast to the standard description of light as a purely transverse wave, longitudinal field components are often significant in cases where the electric field changes significantly on a length scale comparable to $\lambda/(2\pi)$\cite{Bliokh2014}.
In particular, this situation occurs in the case of total internal reflection in the evanescent field. For the case of the quasi linearly polarized nanofiber mode $\rm{HE}^\pm_{11,y}$ (see eq.~\ref{eqnFieldsStart}-\ref{eqnFieldsEnd}), we find a ratio between the longitudinal and transversal field component of
\begin{equation}
\frac{\left|\epsilon_{z}\right|}{\left| \epsilon_{y}\right|}=\frac{2q}{\beta}\frac{\left|\sin(\phi)\right|K_1(qr)}{(1-s)K_0(qr)+(1+s)K_2(qr)\cos(2\phi)}
\label{eqn_long}
\end{equation}
on the surface of the fiber, where the maximum ratio is obtained for $\phi=90^\circ$ and $\phi=270^\circ$. For small radii fibers this ratio increases with the radius and for large radii ($a\gg\lambda$) it approaches $\sqrt{1-\left(n_2/n_1\right)^2}$
for $\phi=90^\circ$ and $\phi=270^\circ$. This expression can also be directly derived for the evanescent field of a  plane wave that undergoes total internal reflection at a planar dielectric interface at grazing incidence\cite{Axelrod84,Kawalec07}. For our experiment we calculate according to eq.~(\ref{eqn_long}) a maximum ratio of $0.557$ which results in an overlap with circular polarization of $93$\%. 

\subsection{Preparation and detection of single nanoparticles on the fiber}
We deposit single gold nanoparticles with a diameter of around $90\,\si{nm}$ on the nanofiber (central resonance wavelength: $\lambda\simeq530\,\si{nm}$, resonant scattering cross section $\sigma_{\textrm{0}}\simeq6\times10^{-3}\,{\rm \mu m^2}$ ). For this purpose, we start with gold nanoparticles dispersed in deionized water (BBI Solutions) and approach the nanofiber with a droplet of this dispersion using a pipette mounted on a three-axis translation stage. We monitor the droplet approaching the nanofiber  with the help of an optical microscope with a long working distance objective (magnification: 20x)  and simultaneously send white light through the fiber to record a real-time absorbance spectra $A(\lambda)=-\log_{10}(I_\textrm{particle}(\lambda)/I_\textrm{ref}(\lambda))$ with a spectrometer. Here $I_\textrm{ref}$ and $I_\textrm{particle}$ are the white light intensity spectra measured before and after the nanoparticle was deposited on the fiber, respectively.
After a successful deposition of a single nanoparticle, a characteristic peak in the absorbance spectrum of about $A_\textrm{max}\simeq0.035$ is observed in the spectrum for our fiber diameter of $315\,\si{nm}$. The height of this peak is proportional to the number of nanoparticles and, together with the shape of the absorbance spectrum,  allows us to determine the number of particles deposited on the fiber. The method yields a positioning accuracy of around 100 $\mu$m in longitudinal direction which is limited by the minimum length over which the droplet touches the fiber. In azimuthal direction, we experimentally obtain an angular uncertainty of $\Delta\Phi\approx20^\circ$. The longitudinal position of the nanoparticle is determined more precisely by sending resonant light through the nanofiber and monitoring the waist region on our camera. In addition, this imaging yields a second estimate of the particle number, and by combining both methods (absorbance and imaging) we can deposit and identify a single nanoparticle on the nanofiber with a success probability close to one. 

In order to measure the nanofiber and nanoparticle diameters as well as to ensure that it is indeed a single particle, we measure the nanofiber waist region with a scanning electron microscope (SEM) after all measurements have been carried out. For this purpose, we deposit the nanofiber on a $2$-inch diameter Al-substrate without rotating the fiber with respect to its initial orientation when the nanoparticle was deposited such that we can measure the azimuthal position of the nanoparticle on the fiber. In order to find the nanoparticle with the SEM on the nanofiber waist (typical length: $l=1\,{\rm cm}$), we deposit -- after the measurements -- groups of additional particles on the fiber at positions $\pm 1,2,3\,\si{mm}$ away from the initial nanoparticle, which serve as markers for finding the correct fiber region. From the measurements, we obtain a diameter of $(315\pm3)\,\si{nm}$ and  $(90\pm3)\,\si{nm}$ for the nanofiber and nanoparticle used in the experiment, respectively. The error bars correspond to the resolution of the SEM. 

\subsection{Modeling the measured data}
The photon fluxes in Fig.~\ref{FigResults2}\textbf{a},\textbf{b} and Fig.~\ref{FigResults1}\textbf{a}-\textbf{d} show the raw data, only corrected for the nonlinearity of the SPCMs. In order to model the measurement results, we calculate the left/right scattering rates for our experimental parameters taking into account the modification of the excitation light field due to the nanofiber\cite{barber90}. According to eq.~(\ref{eqn1}), the photon fluxes are given by
\begin{widetext}
\begin{equation}
c_\pm=\kappa_f \left\{\left|\E_\textrm{exc} \cdot\left[\e^\pm_{\textrm{HE},x}(r,\phi-\phi_0)\right]^*\right|^2+\left|\E_\textrm{exc} \cdot\left[\e^\pm_{\textrm{HE},y}(r,\phi-\phi_0)\right]^*\right|^2\right\}+c_0.
\label{eqnFitFkt}
\end{equation}
\end{widetext}
Here, $\E_\textrm{exc}$ is the excitation light field at the nanoparticle position, that is normalized to one before its modification due to the nanofiber and $\e^\pm_{\textrm{HE}}(r,\phi-\phi_0)$ are the profile functions of the fundamental fiber modes normalized to one on the fiber axis. The parameter $c_{0}=22.5\times 10^3$ $s^{-1}$ accounts for the parasitic coupling of the excitation light into the fiber without nanoparticle (see Fig.~\ref{FigResults2}\textbf{c}), $\phi_0$ is the angular offset of the nanoparticle on the fiber from its expected deposition position at $\phi=90^\circ$ and $\kappa_f$ is the photon flux amplitude.
From a fit of $\kappa_{f}$ and $\phi_{0}$ to the data in Fig.~\ref{FigResults2} we obtain $\phi_{0}=6.3^\circ \pm 0.1^\circ$ and $\kappa_{f}=(21.9 \pm0.1) \times10^6$ s$^{-1}$. Fig.~\ref{FigResults2} shows the results of the fit as well as the theoretical predictions of the simplified model which does not take into account the modification of the light field due to the optical fiber.

From the fitted model function we calculate a polarization averaged cross-section for scattering light into the fiber of $\sigma_f=(5.8\pm2.1)\times 10^{-4}$ $\mu$m$^2$ using our experimental parameters (beam waist at nanoparticle: $w\simeq 150$ $\mu$m, power: $P\simeq 265$ $\mu$W, absolute photon detection efficiency including losses: $\eta\simeq0.46$). Here, the main error contributions originate from uncertainties in the positioning of the beam on the particle and from the estimation of the optical losses in the nanofiber setup.  
In order to compare this value to the theoretical prediction, we start with the resonant free-space scattering cross section $\sigma_{0}\simeq 6\times 10^{-3}$ $\mu$m$^2$ and calculate the fraction scattered into the optical fiber\cite{LeKien06c} for our nanofiber diameter of $315$~nm and an excitation wavelength of $532$~nm. We obtain a polarization averaged cross-section for scattering light into the fiber of $\tilde{\sigma}_f=(7.3\pm1.4) \times 10^{-4}$ $\mu$m$^2$, in good agreement with the measured value. Here, the error is dominated by the uncertainty of the diameter of the nanoparticle.

\section*{Acknowledgements}
\noindent  We gratefully acknowledge financial support by 
the NanoSci-ERA network ``NOIs'' and the European
Commission (IP SIQS, No. 600645). J.V. acknowledges support by the European Commission (Marie Curie IEF Grant 300392).

\section*{Competing Interests}
\noindent The authors declare that they have no
competing financial interests.

\section*{Correspondence}
\noindent  Correspondence and requests for materials
should be addressed to J.V.~(email: jvolz@ati.ac.at) or A.R.~(arno.rauschenbeutel@ati.ac.at). 

\end{document}